\documentclass[preprint2]{aastex}

\usepackage{graphicx}
\usepackage{amsfonts}
\usepackage{times}

\newcommand{\vk}{\mbox{\boldmath $ k $}}
\newcommand{\vA}{\mbox{\boldmath $ A $}}
\newcommand{\vB}{\mbox{\boldmath $ B $}}
\newcommand{\vb}{\mbox{\boldmath $ b $}}

\newcommand{\vF}{\mbox{\boldmath $ F $}}
\newcommand{\vJ}{\mbox{\boldmath $ J $}}

\newcommand{\vU}{\mbox{\boldmath $ U $}}
\newcommand{\vu}{\mbox{\boldmath $ u $}}

\newcommand{\vomega}{\mbox{\boldmath $ \omega $}}
\newcommand{\btimes}{\mbox{\boldmath $ \times $}}

\shorttitle{Hall-MHD Turbulence and Dynamo Action}
\shortauthors{P. D. Mininni, Gomez & Mahajan}

\begin{document}

\title{DIRECT SIMULATIONS OF HELICAL HALL-MHD TURBULENCE AND DYNAMO ACTION}

\author{Pablo D. Mininni\altaffilmark{1}}
\affil{Advanced Study Program, National Center for
       Atmospheric Research, P.O.Box 3000, Boulder CO 80307, USA.}
\and
\author{Daniel O. G\'omez\altaffilmark{2}}
\affil{Departamento de F\'\i sica,
       Facultad de Ciencias Exactas y Naturales,
       Universidad de Buenos Aires, \\
       Ciudad Universitaria, 1428 Buenos Aires, Argentina.}
       \altaffiltext{1}{also at Departamento de F\'\i sica,
       Facultad de Ciencias Exactas y Naturales, Universidad de Buenos Aires,
       Ciudad Universitaria, 1428 Buenos Aires, Argentina.}
       \altaffiltext{2}{also at Instituto de Astronom\'\i a y F\'\i sica del
              Espacio, Ciudad Universitaria, 1428 Buenos Aires, Argentina.}
        \email{dgomez@df.uba.ar}
\and
\author{Swadesh M. Mahajan}
\affil{Institute for Fusion Studies, The University of Texas,
       Austin, Texas 78712, USA.}

\begin{abstract}
Direct numerical simulations of turbulent Hall dynamos are presented. 
The evolution of an initially weak and small scale magnetic field in 
a system maintained in a stationary turbulent regime by a stirring force 
at a macroscopic scale is studied to explore the conditions for 
exponential growth of the magnetic energy. Scaling of 
the dynamo efficiency with the Reynolds numbers is studied, and 
the resulting total energy spectra are found to be 
compatible with a Kolmogorov type law. A faster growth of large scale magnetic 
fields is observed at intermediate intensities of the Hall effect.
\end{abstract}


\section{INTRODUCTION}

In recent years, the relevance of two-fluid effects has been pointed out 
in several astrophysical \citep{Balbus,Sano,MGM1,MGM2} as well as 
laboratory plasmas \citep{Mirnov}. The standard magnetohydrodynamic (MHD)
framework for the study of astrophysical plasmas  may not be adequate in
the presence of strong magnetic fields and/or low ionization;
the electric conductivity then is not isotropic and nonlinear effects 
arise in Ohm's law.  In low temperature accretion disks around young 
stellar objects or in dwarf nova systems in quiescence, for example, the plasma is 
only partially ionized with a small abundance of charged particles. As a 
result, two new effects appear in Ohms's law: ambipolar diffusion and Hall 
currents. The predominance of either of these effects is determined by 
the ionization fraction and the plasma
density \citep{Sano,Braginskii}. The impact of ambipolar diffusion on 
dynamo action has been already studied by \citet{Zweibel} (see also 
\citet{BraSub}). The Hall effect 
affects the dynamics of protostellar disks \citep{Balbus}. In neutron stars 
magnetic fields are so strong that the Hall term can be even more important 
than the induction term for the magnetic field evolution \citep{Muslimov}. 
The Hall effect is also known to be relevant in others astrophysical scenarios 
\citep[and references therein]{Sano,MGM3}.

The Hall effect, when strong, is expected to seriously affect the MHD results
on the  generation of magnetic fields in astrophysical and 
laboratory plasmas by inductive motions in a conducting fluid (dynamo 
effect). More specifically, it is expected to modify the growth and evolution 
of magnetic energy, since the addition  of the Hall term to the MHD equations 
leads to the freezing 
of the magnetic field to the electron flow (in the non-dissipative limit) 
rather than to the bulk velocity field. The first studies on the impact of 
Hall currents on dynamo action \citep{Helmis1,Helmis2} were carried out
using mean field theory and the first-order smoothing approximation 
\citep{Krause}. Helmis obtained decreasing dynamo action as the strength 
of the Hall terms increased. Recently, the impact of the Hall effect 
has been studied using the kinematic approximation or focusing on particular 
geometries, both in numerical simulations \citep{Galanti} and in experimental 
and theoretical 
studies \citep{Heintzmann,Hantao,Rheinhardt,Mirnov}. A general closure scheme
was also proposed to compute the contribution of the Hall term to the dynamo action
$\alpha $-effect \citep{MGM1} using mean field theory and the reduced 
smoothing approximation \citep{Blackman}; it was found that  the Hall effect 
could either suppress or enhance dynamo action. Expressions for the 
turbulent diffusivity using this closure were also derived for particular 
cases in \citet{MGM2}.

In the present work, we report results from direct numerical simulations of 
the dynamo action in MHD and Hall-MHD at moderate Reynolds numbers  with strong 
kinetic helical forcing. The study of the scaling of the dynamo 
efficiency with increasing Reynolds numbers is the main aim of this paper. 
In a previous paper \citep{MGM3}, we showed that three distinct dynamo 
regimes can be clearly identified: (1)  dynamo activity is 
enhanced (Hall-enhanced regime), (2) it is inhibited (Hall-suppressed regime), 
(3) it asymptotically approaches the MHD value (MHD regime). These 
regimes arise as a result of the relative ordering between the relevant 
lengthscales of the problem. Namely, the energy-containing scale of the 
flow, the Hall length, and the correlation length of the magnetic seed.
Simulations in \citet{MGM3} were performed with $64^3$ spatial grid points 
and fixed Reynolds numbers.  This study left unanswered the  question whether 
the enhancement of dynamo action by Hall effect  would increase or decrease 
with increasing Reynolds numbers and scale separation.

Theoretical estimates using mean field theory and a particular choice 
for the small scale fields suggest that the efficiency of Hall-MHD dynamos 
compared with the MHD counterpart  will  increase as the scale separation 
is increased \citep{MGM1}. In this work we present direct simulations with 
higher spatial resolutions and Reynold numbers that confirm this result. 
Also, the magnetic, kinetic, and total energy spectra developed in Hall-MHD 
turbulence are calculated and studied. These spectra correspond to the dynamo regime, 
where no imposed currents are present, i.e, the  magnetic fields are purely 
self-generated. This regime is rather 
general and probably common to most astrophysical flows \citep{Haugen04}.

The paper is organized as follows. In Section 2 we present the general 
equations describing the evolution of the fields. In Section 3, the code 
used to numerically integrate the Hall-MHD system is described, and all 
the simulations made with different resolutions and parameters are listed. 
Section 4 presents the results obtained in the MHD limit of our equations; the results 
obtained are similar to those of other authors \citep{Meneguzzi,Brandenburg}, 
and provide 
a reference  set to compare the Hall-MHD solutions with. Section 5 is devoted to 
the results of Hall-MHD simulations. In Section 6 we discuss a subset of 
MHD and Hall-MHD simulations made to study the evolution of large scale 
magnetic fields. Finally, in Section 7, we give a brief summary of the 
current effort.

\section{THE HALL-MHD SYSTEM}

Incompressible Hall-MHD is described by the modified induction and the 
dissipative Navier-Stokes equation,
\begin{eqnarray}
\frac{\partial\vB}{\partial t} & = & \nabla\btimes\left[\left( \vU
     - \epsilon\nabla\btimes\vB\right)\btimes\vB\right] 
     + \eta \nabla^2 \vB \label{HallMHD} \\
\frac{\partial \vU}{\partial t} & = & - \left( \vU \cdot \nabla \right) \vU  
     + \left( \vB \cdot \nabla \right) \vB - {} \nonumber \\
 & & {} - \nabla \left( P + \frac{B^2}{2} \right) + \vF + \nu \nabla^2 \vU 
\label{NS}\\
\nabla\cdot\vB & = & 0\ =\ \nabla\cdot\vU \; ,\label{divUB}
\end{eqnarray}
where $\vF$ denotes a solenoidal external force. The velocity $\vU$ 
and the magnetic field $\vB$ are expressed in units of a characteristic speed 
$ U_0 $, $ \epsilon $ measures the relative strength of the Hall effect, and  
$ \eta $ and $\nu$ are the (dimensionless)  magnetic diffusivity and kinematic 
viscosity, respectively. Note that the measure of Hall effect  $ \epsilon $ can 
be written as
\begin{equation}
\epsilon = \frac{L_{Hall}}{L_0} \; ,
\end{equation}
where $ L_0 $ is a characteristic length scale (the size of the box in our 
simulations is equal to $2\pi$), and the Hall length
\begin{equation}
L_{Hall} = \frac{c}{\omega_{pi}}\ \frac{U_A}{U_0} \; ,
\label{LHall}
\end{equation}
is given in  terms of the characteristic speed $ U_0 $ and the characteristic 
Alfv\'enic speed $ U_A $. In particular, we are free to choose 
$U_0 = U_A $ as our characteristic velocity, reducing $L_{Hall}$ 
to the ion skin depth. 

Equation (\ref{LHall}) is valid in a fully ionized plasma. We will work 
under this assumption without any loss of generality.
In the more general case of partially ionized plasmas, the 
values of $ L_{Hall} $ and $ \epsilon $ are those reported by 
\citet{Sano}, or in \citet{MGM3}. Typical values of $\epsilon$ in 
astrophysics are also mentioned in these papers. 

To keep in mind astrophysical scenarios, we just recall three examples. 
In a protostellar disk the Hall scale is larger than the dissipation scale 
typically by two orders of magnitude, but smaller than the largest scales 
of the system \citep{Balbus}. Therefore we expect $\epsilon<1$ but with 
$L_{Hall}$ larger than the dissipation scale. In some dwarf nova disks 
and protoplanetary disks $\epsilon \approx 1 $ \citep{Sano}. As previously 
mentioned, in neutron stars  $\epsilon>1$ \citep{Muslimov}.

The Hall-MHD system has three well-known ideal ($ \eta = \nu = 0 $) 
quadratic invariants 
\begin{eqnarray}
E & = & \frac{1}{2} \int (U^2 + B^2) \, dV \; , \\
H_m & = & \frac{1}{2} \int \vA \cdot \vB \, dV \; , \\
K & = & \frac{1}{2} \int (\vB + \epsilon \vomega) \cdot 
     (\vA + \epsilon \vU)\, dV \; .
\end{eqnarray}
Here $ E $ is the energy, $ H_m $ is the magnetic helicity, and $ K $ is the 
hybrid helicity, which replaces the cross helicity from magnetohydrodynamics. 
The vector potential $ \vA $ is defined by $ \vB = \nabla \times \vA $, and 
$\vomega=\nabla \times \vU$ is the vorticity. Conservation of these ideal 
invariants during the evolution of the system provides a check on the 
simulation.

\section{THE CODE}

The pseudospectral code used in \citet{MGM3} was modified to run in a 
Beowulf cluster using MPI. We integrated the Hall-MHD equations 
(\ref{HallMHD})-(\ref{divUB}) in a cubic box with periodic boundary conditions. 
The equations were evolved in time using a second order Runge-Kutta method.
The total pressure $ P_T = P + B^2/2 $ was computed in a self-consistent 
fashion at each time step to ensure the incompressibility condition 
$ \nabla \cdot \vU = 0 $ \citep{Canuto}. In Fourier space, taking the 
divergence of equation (\ref{NS}) we obtain
\begin{equation}
\widehat{P_T}(\vk) = \frac{i}{k^2} \vk \cdot \left[\widehat{\left(\vU \cdot \nabla 
     \vU \right)_{\vk}} -\widehat{\left(\vB \cdot \nabla \vB \right)_{\vk}}\right] \; ,
\label{pres}
\end{equation}
where the hat denotes a spatial Fourier transform, and $\vk$ is the wavenumber 
vector.

To satisfy the divergence-free condition for the magnetic field, the 
induction equation (\ref{HallMHD}) was replaced by an equation for the 
vector potential
\begin{equation}
\frac{\partial \vA}{\partial t} = \left( \vU - \epsilon \nabla \times \vB 
     \right) \btimes \vB + \epsilon\nabla p_e + \eta \nabla^2 \vA \; ,
\label{vecpot}
\end{equation}
where $ p_e $ (electron pressure) was computed at each time step to satisfy 
the Coulomb gauge $ \nabla \cdot \vA = 0 $, solving an equation similar to equation 
(\ref{pres}).

We present results from different runs with $\eta = 0.05$, 
$\eta = 0.02$, and $\eta = 0.011$. For the first value of  
$\eta$, simulations with $64^3$ and $128^3$ spatial grid points were 
performed to check convergence. The rest of the simulations were made with 
$128^3$ grid points ($ \eta = 0.02$), and $256^3$ grid points 
($\eta = 0.011$). All the runs were made with magnetic Prandtl 
number $\nu/\eta=1$. Therefore, hereafter we will only consider a 
single Reynolds number (i.e. both kinetic and magnetic), defined as
\begin{equation}
R = \frac{UL_0}{\eta} \; ,
\end{equation}
In the study of turbulent flows, the number
\begin{equation}
R_\lambda = \frac{U\lambda}{\eta} \; ,
\end{equation}
constructed from Taylor's length scale 
[$\lambda = (\left<U^2\right>/\left<\omega^2\right>)^{1/2}$] is often
considered. Note that this definition of Taylor's micro-scale might differ 
from other definitions (for instance, in connection with experiments on 
fluid turbulence) in factors of order unity \citep{Pope}. These two Reynolds 
numbers were respectively $R \approx 100$ and 
$R_\lambda \approx 20$ for the first value of $\nu$ and $\eta$, 
$R \approx 300$ and $R_\lambda \approx 40$ for the second one, and 
$R \approx 560$ and $R_\lambda \approx 60 $ in the last case. The 
energy injection rate was approximately the same for all these runs.

The simulation begins by subjecting the Navier-Stokes 
equation to a stationary helical force $\vF$ (given by eigenfunctions of the 
curl operator) operating at a macroscopic scale $k_{force}=3$
 \citep{MGM3} to reach a hydrodynamic turbulent steady state.  The resulting 
 statistically steady state is characterized by a positive kinetic helicity. 
The relative helicity in runs with $R=300$ is 
$2\ H_k/(\left<U^2\right>\left<\omega^2\right>)^{1/2} \approx 0.4 $, and this value 
decreases slightly for larger Reynolds numbers. The kinetic helicity 
is defined as 
\begin{equation}
H_k = \frac{1}{2} \int \vU \cdot \vomega \, dV \; .
\end{equation}

Once the hydrodynamic stage of the simulation reaches a steady state, a non-helical 
but small magnetic seed was introduced. This initial magnetic seed was generated 
by a $ \delta $-correlated vector potential centered at $ k_{seed} = 13 $ for 
the $R = 100$ runs and $ k_{seed} = 35 $ for the $R = 300$ and $R = 560$ 
simulations. The run was continued with the same external helical force in the 
Navier-Stokes equation, to study the growth of magnetic energy due to dynamo action. 

Another set of simulations was made under the same conditions but with 
$\nu = \eta = 0.02$ and $k_{force}=10$ ($128^3$ grid points), to study 
the changes in the growth of the large scale magnetic field in the presence of 
the Hall effect. In this case the Reynolds numbers are 
smaller ($R\approx220$ and $R_\lambda\approx20$), and the turbulence is 
weaker, since there are not enough modes in Fourier space for a direct 
cascade to develop properly. On the other hand, there are more Fourier 
modes to study the inverse cascade and the growth of large scale fields. 
The results of these simulations are discussed in Section 6.

In all our simulations the Kolmogorov's kinetic and magnetic dissipation 
length scales were properly resolved in the computational domain, i.e. we 
made sure that the dissipation wavenumbers remain smaller than the maximum 
wavenumber allowed by the dealiasing step, namely $k_{max}= 128/3$.

\section{MHD DYNAMOS}

In this section we briefly present the results from MHD simulations. The 
results are in good agreement with previous simulations of dynamo action 
under periodic boundary conditions \citep{Meneguzzi,Brandenburg}. These 
simulations are intended for comparison with the Hall-MHD runs, and 
therefore some specific results of the MHD simulations are discussed in 
more detail in the following sub-sections.

\subsection{Magnetic energy evolution}

\begin{figure}
\plotone{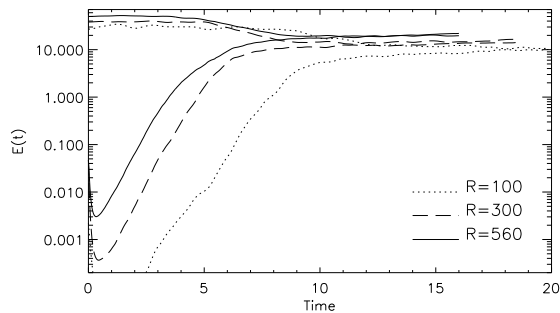}
\caption{Kinetic energy $E_k$ (above), and magnetic energy $E_m$ (below) 
         as a function of time ($ \epsilon = 0 $).
\label{mhd_ener}}
\end{figure}

In Figure \ref{mhd_ener}, we show the magnetic and kinetic energy as a 
function of time in MHD runs ($ \epsilon = 0 $) for different Reynolds 
numbers. The turnover time for  $R=300$  is
$\tau=2\pi /(k_{force} \left<U^2\right>^{1/2}) \approx 0.3$. 

Two phases can be clearly identified in the evolution of the magnetic energy. 
After a first stage with exponential growth (which can be considered 
as the  kinematic dynamo stage), the magnetic energy saturates and reaches 
equipartition with the kinetic energy (see Figure \ref{mhd_ener}). In the 
first stage, the magnetic energy is still weak and the velocity field is 
not strongly affected by the Lorentz force. Note that during this 
exponential growth of magnetic energy, the kinetic energy remains 
approximately constant.

This kinematic dynamo stage can be understood at least qualitatively 
considering the mean field induction equation \citep{Krause}
\begin{equation}
\frac{\partial \overline{\vB}}{\partial t} = 
     \nabla \btimes \left( \overline{\vU} \btimes \overline{\vB} + 
     \alpha \overline{\vB} \right) + \eta_{ef\!f} \nabla^2 \overline{\vB} .
\label{MFMHD}
\end{equation}
Here the overline denotes mean field quantities, and $\eta_{ef\!f}$ 
is the magnetic plus turbulent diffusivity. The MHD $\alpha$-effect 
\citep{Pouquet}
\begin{equation}
\alpha = \frac{\tau}{3} \left( - \overline{\vu \cdot \nabla \btimes
     \vu} + \overline{\vb \cdot \nabla \btimes \vb} \right) ,
\label{alphaMHD}
\end{equation}
represents the back-reaction of the turbulent motions in the mean field, 
and gives exponential growth of magnetic energy in the kinematic regime 
for helical turbulence. Here $ \vu $ and $ \vb $ are respectively the 
fluctuating velocity and magnetic fields, and $\tau$ is a typical 
correlation time for the turbulent motions. Attempts to measure 
this quantity in direct simulations were made by \citet{Cattaneo}, and 
\citet{Brandenburg}.

As the Reynolds number increases, the saturation field strength increases, 
although it seems to reach an asymptotic value.
After the saturation, the magnetic energy keeps growing slowly on a 
resistive timescale \citep{Brandenburg}. This late growth takes place 
mainly at large scales, as will be shown in the energy spectrum. As 
far as we know, there are no simulations of MHD dynamos with periodic 
boundary conditions showing generation of large scale fields on 
shorter times.

\subsection{Energy spectrum}

\begin{figure}
\plotone{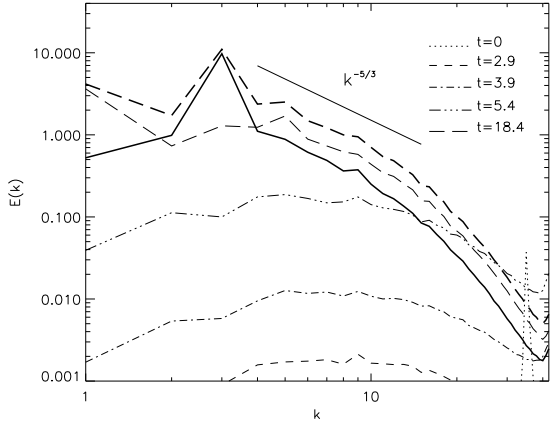}
\caption{Mean kinetic energy spectrum (thick line), total energy 
         spectrum (thick dashed line), and magnetic energy spectrum 
         at different times ($\epsilon =0$ and $R=300$). The 
         Kolmogorov's slope is shown as a reference.
\label{mhd_spec}}
\end{figure}

Figure \ref{mhd_spec} shows the kinetic and magnetic spectra at different 
times for a run with $ R = 300 $. In the early stages the magnetic energy 
grows uniformly at all wave numbers. After the saturation ($t\approx5$) the 
emergence of a large-scale field can be clearly seen in the spectrum.  
At $t\approx18.4$,  when the system has already
reached equipartition (see Figure \ref{mhd_ener}), the magnetic 
energy at large scales (small wave numbers) still keeps growing, albeit slowly. As a 
result, the large scale magnetic field reaches super-equipartition with the 
kinetic energy. An excess of magnetic energy can be also observed at small 
scales. 

The slope of the total (magnetic and kinetic) energy spectrum in the 
inertial range is consistent with Kolmogorov's $k^{-5/3}$ law and in 
good agreement with simulations of helical MHD turbulence with higher 
spatial resolutions. \citet{Kida} found that the total energy spectrum 
in MHD dynamo simulations is of the form
\begin{equation}
E(k) = C_K \varepsilon^{2/3} k^{-5/3} ,
\end{equation}
where $\varepsilon$ is the total dissipation rate, and $C_K$ is a Kolmogorov's 
constant. Simulations at higher spatial resolutions \citep{Haugen03} seem to 
confirm this result although in some cases the spectrum tends to be a 
little shallower.

\subsection{Magnetic helicity}

\begin{figure}
\plotone{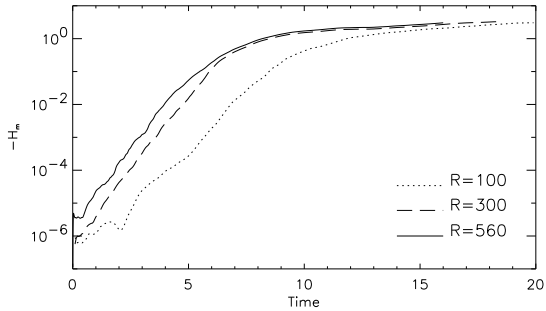}
\caption{Magnetic helicity ($ -H_m $).
\label{mhd_hel}}
\end{figure}

The magnetic helicity $ H_m $ is displayed in Figure \ref{mhd_hel}. The 
initial magnetic field is non-helical, but during the dynamo process 
net magnetic helicity is generated with a sign opposite to that of the 
kinetic helicity. This helicity is located mostly in large scale structures, 
as will be shown in Section 6. 

From mean field equations we obtain for the large scale magnetic helicity 
\citep{MGM3}
\begin{equation}
\frac{d \overline{H_m}}{dt} = 2 \int \left(\alpha \overline{B}^2 - 
     \eta_{ef\!f} \overline{\vJ} \cdot \overline{\vB}\right) \, dV \; ,
\label{htransfer}
\end{equation}
where $\vJ=\nabla \times \vB$ is the electric current density. This equation 
represents a transfer of magnetic helicity from small scales to large 
scales. The mean field helicity grows with the same sign as the $ \alpha $ 
coefficient (opposite sign as the kinetic helicity). Our results are in 
good agreement with this relation, as well as previous simulations of MHD 
dynamo action \citep{Brandenburg,MGM3}.

Although this generation of magnetic helicity by dynamo action is expected 
to decrease as the magnetic Reynolds number $ R_m $ increases, in 
simulations with higher Reynolds numbers the growth of magnetic helicity 
seems to reach an asymptotic value.

\section{HALL DYNAMOS}

To quantitatively assess the role of the Hall effect on dynamo action, 
we display results from runs with different values of  $\epsilon$ 
: the MHD run ($ \epsilon = 0 $) and Hall-MHD runs with 
$ \epsilon = 0.066, 0.1$ and $0.2$. We will focus on the 
Hall-enhanced dynamo regime \citep{MGM3}. Note that all these 
values of $ \epsilon $ correspond to dynamos where the Hall effect is 
only relevant in a fraction of the scales involved. The Hall inverse 
length scale for these runs is measured by $ k_{Hall} = 15, 10$ and 5 
respectively ($k_{Hall}=1/\epsilon$). All length scales smaller than the 
Hall scale are expected to be strongly affected by the Hall effect. The 
Kolmogorov's kinetic dissipation scale 
[$ k_\nu = ( \left< \omega^2 \right>/ \nu^2 )^{1/4}$] is 
$ k_\nu \approx 20 $ when $ R=100 $, $ k_\nu \approx 40 $ when 
$ R=300 $, and $ k_\nu \approx 75 $ when $ R=560 $. 

\subsection{Magnetic energy evolution}

\begin{figure}
\plotone{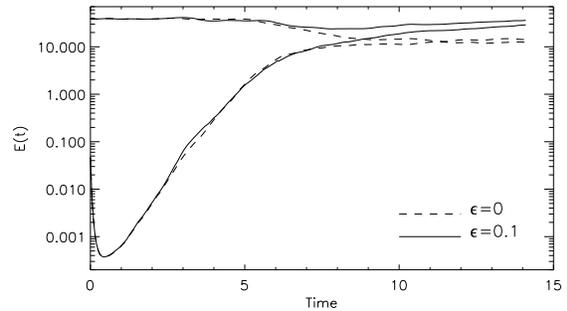}
\caption{Magnetic (below) and kinetic energy (above) as a function 
         of time for two runs with $ \epsilon = 0.1$ and $ \epsilon = 0$ ($R=300$).
\label{hall1_ener}}
\end{figure}

Figure \ref{hall1_ener} shows the kinetic and magnetic energy as a 
function of time for the MHD and a Hall-MHD run with $ R = 300 $ and 
$ \epsilon = 0.1 $. At early times, the evolution of magnetic energy in 
MHD and Hall-MHD is similar.
Using mean field theory, the induction 
equation for the mean magnetic field reduces to 
\begin{equation}
\frac{\partial \overline{\vB}}{\partial t} = 
     \nabla \btimes \left[ \left(\overline{\vU} - \epsilon \nabla \btimes 
     \overline{\vB} \right) \btimes \overline{\vB} + 
     \alpha \overline{\vB} \right] + \eta_{ef\!f} \nabla^2 \overline{\vB} ,
\label{MFHALL}
\end{equation}
with the Hall-MHD $\alpha$-effect now given by \citep{MGM1}
\begin{eqnarray}
\alpha & = & \frac{\tau}{3} \left( - \overline{\vu^e \cdot \nabla \btimes
     \vu^e} + \overline{\vb \cdot \nabla \btimes \vb} 
     - {} \nonumber \right. \\
& & {} \left. - \epsilon \, \overline{\vb \cdot \nabla \btimes \nabla \btimes 
     \vu^e} \right) \; .
\end{eqnarray}
Here $ \vu^e = \vu - \epsilon \nabla \btimes \vb $ is the fluctuating 
electron flow velocity. When the fluctuating magnetic field is 
weak, this expression reduces to equation (\ref{alphaMHD}) and the Hall 
effect can be dropped. Therefore, the first stage corresponds to a kinematic 
dynamo during which the Hall effect is negligible.

\begin{figure}
\plotone{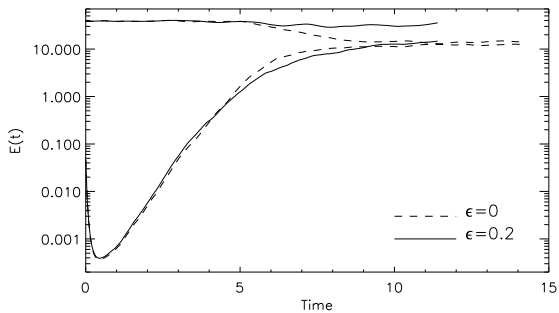}
\caption{Magnetic (below) and kinetic energy (above) as a function 
         of time for two runs with $ \epsilon = 0.2 $ and 0 ($R=300$).
\label{hall2_ener}}
\end{figure}

After this stage, and when the dynamo-generated magnetic fields are strong 
enough for the Hall effect to become non-negligible, the evolution changes and the 
magnetic energy keeps growing but at a different pace (see also Figure 
\ref{hall2_ener}). A third stage can be identified, when the velocity 
field is affected by the Lorentz force and the dynamo reaches saturation. The 
kinetic energy drop is not as intense as in the MHD case, and the 
increase of magnetic energy in this final stage is larger than in the 
MHD case for moderate values of $\epsilon$. Finally, a state with more 
magnetic energy than its MHD counterpart is reached (by a factor $2.3$ 
when $R=300$ and $\epsilon=0.1$).

Figure \ref{hall2_ener} shows the evolution for a Hall-MHD run with 
$ R = 300 $ and $ \epsilon = 0.2 $. Here, after the exponentially growing 
stage, the magnetic field saturates at an amplitude smaller than the 
previous run. Note that equipartition between the kinetic and magnetic 
energy is not reached.

When a turbulent stationary state is attained, the sum of magnetic 
and kinetic energy is not equal to the initial kinetic energy. This 
is related to the fact that, when the magnetic seed is introduced, a new 
channel for energy dissipation arises,
\begin{equation}
\frac{dE}{dt} = - \nu \int \omega^2 \, dV - \eta \int J^2 \, dV \; ,
\end{equation}
As in previous simulations \citep{MGM3}, it is found that the final energy 
reached for the Hall-MHD runs is larger than the value obtained for the 
MHD runs, revealing that Hall-MHD dynamos can be more efficient (in the 
sense that they generate more magnetic energy and dissipate less total 
energy).

\begin{figure}
\plotone{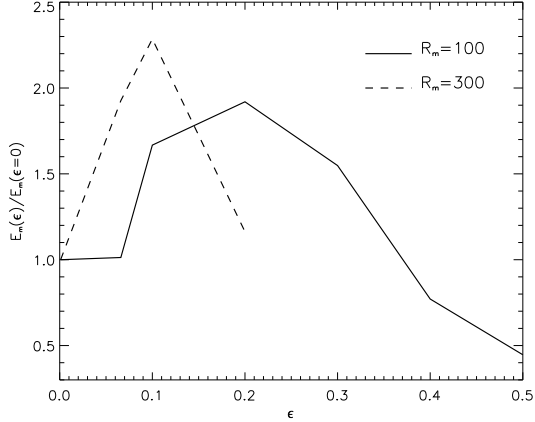}
\caption{$E_m/E_m(\epsilon=0)$ as a function of $\epsilon$ for runs 
         with $R=100$ and $R=300$. The values of $E_m$ correspond to 
the magnetic energy at the saturation level, while $E_m(\epsilon=0)$ 
is the magnetic energy for an MHD run
\label{hall_runs}}
\end{figure}

Figure \ref{hall_runs} shows the maximum value attained by the magnetic 
energy as a function of $\epsilon$ for several simulations with different 
Reynolds numbers and scale separations. When $R=560$ the final amplitude reached by the 
magnetic energy is unknown. Given the stringent quadratic Courant-Friedrich-Lewy (CFL) 
condition imposed by dispersive waves in Hall-MHD, simulations were only carried up to 
saturation of the dynamo. The maximum value of the energy in Figure 
\ref{hall_runs} is normalized with the value obtained in an MHD run with 
the same Reynolds numbers and initial kinetic energy and helicity. 
As previously mentioned, considering the three different regimes of the 
Hall dynamo discussed in 
\citet{MGM3}, we focus on the Hall-enhanced case. As the Reynolds numbers 
are increased, the efficiency of the Hall-MHD dynamo grows. Also, the value of 
$\epsilon$ at which maximum efficiency is obtained, decreases as the Reynolds numbers 
are increased. Note that the growth of the efficiency of 
the Hall-MHD dynamo with increasing scale separation was predicted analytically by 
\citet{MGM1}. The shift of most efficient 
$\epsilon$ to smaller values as $R$ increases is also obtained from 
analytical estimates \citep{MGM4}.

\subsection{Energy spectrum}

\begin{figure}
\plotone{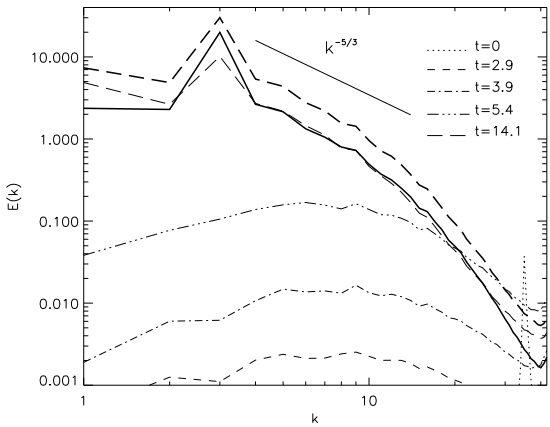}
\caption{Mean kinetic energy spectrum (thick line), total energy 
         spectrum (thick dashed line), and magnetic energy spectrum 
         at different times ($\epsilon = 0.1$ and $R=300$).
\label{hall_spec}}
\end{figure}

Figure \ref{hall_spec} shows the kinetic, magnetic, and total energy 
spectra at different times for $ R = 300 $ and $ \epsilon = 0.1 $. 
Barring $\epsilon$, all the parameters and initial 
conditions in this run are the same as those corresponding to Figure 
\ref{mhd_spec}. Therefore, a direct comparison between the evolution of 
both spectra can be made. During the first few time steps, the evolution is 
similar to the MHD run, with the entire magnetic spectrum growing at 
almost the same rate. The difference observed in \citet{MGM3}, that the 
large-scale magnetic field is slightly larger than in its MHD counterpart, 
is now increased as a result of the larger Reynolds number and larger 
scale separation. 

\begin{figure}
\plotone{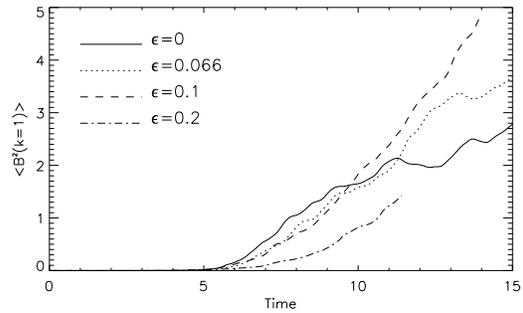}
\caption{Magnetic energy in the shell $k=1$ in Fourier space as a 
         function of $\epsilon$ and time.
\label{Bmean}}
\end{figure}

The rate of increase of the large scale magnetic field changes in the presence of 
the Hall effect. While in MHD the build-up of this field proceeds on 
a resistive timescale, in Hall-MHD it  grows faster. Note that the 
magnetic energy in the shell $k=1$ in Hall-MHD ($\epsilon=0.1$) 
simulation at $t=14.1$ is a factor of 2 larger than the magnetic 
energy in the same shell in the MHD run at $t=18.4$. Also, the kinetic 
energy in the same shell is larger. This can be 
also observed in Figure \ref{Bmean}, which shows the energy contained 
in the large scale magnetic field (in the shell $k=1$) as a function 
of time for several values of $\epsilon$. However, the filling factor 
$(\left<\overline{B}^2\right>/\left<B^2\right>)^{1/2}$ is smaller as 
$\epsilon$ is increased. At $t=14$ the filling factor is $\approx 0.23$ 
for $\epsilon=0$, $\approx 0.17$ for $\epsilon=0.1$, and $\approx 0.1$ 
for $\epsilon=0.2$, 

Note also that, while the MHD spectrum shows super-equipartition at small 
scales (the magnetic energy is larger than the kinetic energy at large 
wave numbers),  the Hall-MHD leads to equipartition at these
scales. Both the evolution of the large scale and small scale magnetic 
fields are therefore clearly affected by the Hall effect, even though the 
Hall effect operates effectively  only at small scales.

\begin{figure}
\plotone{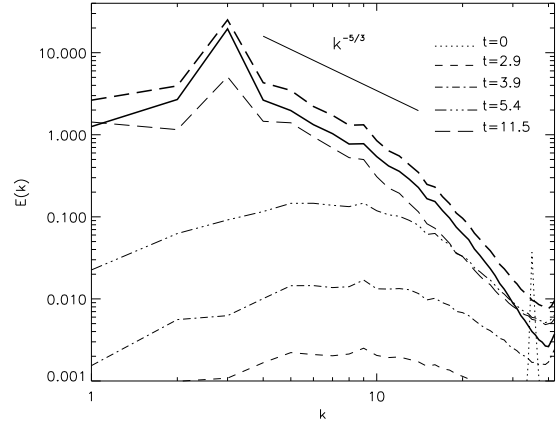}
\caption{Mean kinetic energy spectrum (thick line), total energy 
         spectrum (thick dashed line), and magnetic energy spectrum 
         at different times ($\epsilon = 0.2$ and $R=300$).
\label{hall2_spec}}
\end{figure}

Figure \ref{hall2_spec} shows the spectrum for $R=300$ and $\epsilon=0.2$. 
In this case all wave numbers larger than $k_{Hall}=5$ are affected by 
the Hall effect. However, the total energy spectrum in the saturated state 
seems to obey a Kolmogorov type law, although the Hall length scale is placed 
in the middle of the inertial range.

\begin{figure}
\plotone{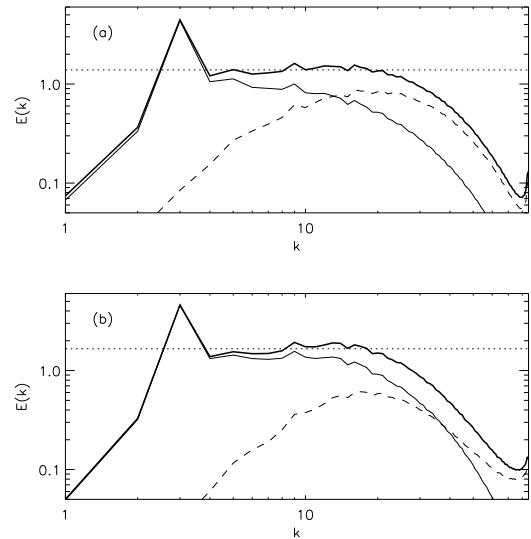}
\caption{Total compensated energy spectrum $k^{5/3}\varepsilon^{-2/3}E_k$ 
         (thick line), kinetic energy 
         spectrum (thin line), and magnetic energy spectrum (dashed line) at 
         $t=5.5$ for a run with (a) $\epsilon = 0$, and (b) $\epsilon = 0.1$ 
         ($R=560$, $256^3$ grid points).
\label{hall_spec2}}
\end{figure}

Figure \ref{hall_spec2} shows the compensated energy spectrum 
$E(k) / (\varepsilon^{2/3} k^{-5/3})$ for higher spatial resolution runs 
with $ R = 560 $ and $ \epsilon = 0$ and $0.1$, using $256^3$ grid points. 
If the spectrum obeys a Kolmogorov  type law, the compensated spectrum should 
be flat over a certain range, and the amplitude of the spectrum in this 
range gives the Kolmogorov constant $C_K$. The mild hump observed for the total 
energy before entering the dissipative range, might be indicative of the presence 
of a ``bottleneck effect" for the energy cascade, as was recently discussed by 
\citet{Haugen03}.

The spectra displayed in Figure \ref{hall_spec2} correspond  to the era when the   
dynamo is saturated. The first example (Fig.~\ref{hall_spec2}(a)) represents the 
MHD limit (i.e. $\epsilon = 0$), while 
the second one (Fig.~\ref{hall_spec2}(b)) has $k_{Hall}=10$ ($\epsilon = 0.1$). 
All the length scales smaller 
than $1/k_{Hall}$ are expected to be dominated by the Hall 
effect. For the  MHD spectrum, the Kolmogorov constant  $C_K\approx1.39$, a value slightly 
larger than the one found by \citet{Haugen03}, $C_K\approx1.3$ (see also \citet{Haugen04}), but 
smaller than the one obtained by \citet{Kida} ($C_K\approx2.1$). No clear change 
in the slope can be identified in the Hall-MHD case, and the spectrum is 
compatible with a Kolmogorov type law in its inertial range. However, the 
spectrum could really be  a little shallower. Higher spatial resolution
simulations are needed to settle this point. In this case the 
Kolmogorov's constant turns out to be $C_K\approx1.66$, somewhat larger than its 
MHD counterpart. The Kolmogorov's dissipation wave numbers in the last 
stages of these runs are 
$k_\nu \approx k_\eta \approx 70$ 
[where $k_\eta = (\left<J^2\right>/\eta^2)^{1/4}$].

\subsection{Magnetic helicity}

\begin{figure}
\plotone{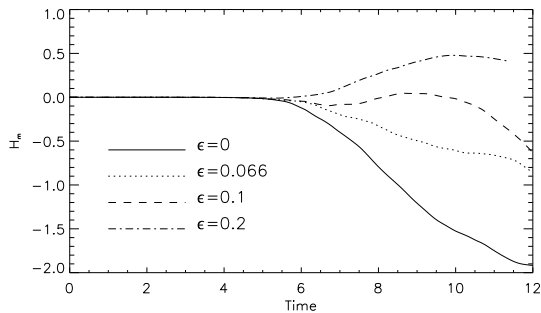}
\caption{Magnetic helicity for $ \epsilon = 0, 0.066, 0.1 $, and $0.2$ 
         ($R=300$).
\label{hall_helm}}
\end{figure}

In \citet{MGM3}, the Hall effect was observed to inhibit the creation of 
net magnetic helicity by the dynamo process. This effect is enhanced as 
we increase the Reynolds numbers. While the MHD dynamo is an efficient 
generator of magnetic helicity with most of this helicity concentrated 
in the larger scales,  the Hall dynamo is somewhat sluggish;  the growth of 
net magnetic helicity is slower and in some cases oscillates around zero 
(see Figure \ref{hall_helm}.) This result is in good agreement with 
theoretical estimates suggesting that in the presence of the Hall effect 
reconnection events are faster \citep{Priest,Hantao}, and therefore 
dissipate less magnetic helicity (see Section 6).

\subsection{Kinetic helicity}

In MHD, relative kinetic helicity is known to change only slightly 
during dynamo action \citep{Brandenburg}. In all these simulations, 
kinetic energy and kinetic helicity are injected at the same length scale 
by the stirring force acting at $k_{force}=3$. For homogeneous hydrodynamic 
turbulence, kinetic helicity directly cascades to smaller scales. On 
dimensional grounds \citep{Moffat}, the spectrum of kinetic helicity in 
the inertial range is also expected to follow Kolmogorov's law 
\citep{Eyink,Gom2004}
\begin{equation}
H_k(k) = C_H k_{hel} \varepsilon^{2/3} k^{-5/3} \; ,
\end{equation}
where $C_H$ is another  Kolmogorov constant, and $k_{hel}$ is the scale where 
kinetic helicity is injected. Using the inertial range kinetic energy expression ($C_K$ is the standard Kolmogorov's constant )  
\begin{equation}
E_k(k) = C_K \varepsilon^{2/3} k^{-5/3} \; ,
\end{equation}
the ratio between the 
total kinetic helicity and kinetic energy in hydrodynamic turbulence come out to be
\begin{equation}
\frac{H_k}{E_k} = \frac{\int{\vU \cdot \vomega \, dV}}
    {\int{U^2 dV}} \approx k_{hel} \; ,
\end{equation}
where $k_{hel}=k_{force}$ in our case. What happens after we introduce the 
magnetic seed?

\begin{figure}
\plotone{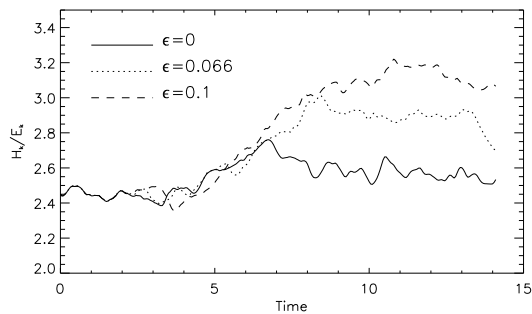}
\caption{$H_k/E_k$ for $ \epsilon = 0, 0.066$, and $0.1$.
\label{hall_helk}}
\end{figure}

Figure \ref{hall_helk} shows the ratio $H_k/E_k$ for the MHD and the 
Hall-MHD runs with $R=300$. The evolution of the relative kinetic helicity 
is similar. In the MHD run $H_k/E_k \approx 2.5$, a value close to 
$k_{force}=3$. Note that, although the total kinetic energy (and the 
kinetic helicity) decreases during the time evolution as a result of the 
increasing Lorentz force, the ratio $H_k/E_k$ remains nearly constant. 
On the other hand, this ratio grows with $\epsilon$ in the Hall-MHD runs. 
The growth takes place just after the exponentially growing stage, when 
a large scale magnetic field is developing in the box. This result 
suggests that a new source of kinetic helicity has appeared, and is in 
good agreement with theoretical estimates suggesting that the Hall effect 
introduces handedness in the fluid motions \citep{MGM4}. However, we want 
to point out that this handedness does not, by itself, generate a net $\alpha$-effect.

As previously mentioned, the relative kinetic helicity 
$H_k/(\left<U^2\right>\left<\omega^2\right>)$ in   Hall-MHD  also 
shows the same behavior, changing from a relative kinetic helicity 
of $0.4$ when $\epsilon = 0$ up to about $0.6$ when $\epsilon = 0.1$.

\section{LARGE SCALE MAGNETIC FIELD GENERATION}

In this section we present MHD and Hall-MHD results for an external 
force located at $k_{force}=10$. Simulations were carried with $\nu=\eta=0.02$ 
and $128^3$ grid points. In this case the Reynolds numbers are smaller 
($R\approx220$ and $R_\lambda\approx20$), and the turbulence is weaker, 
since there are not enough modes in Fourier space for a direct cascade 
to develop properly. On the other hand, there is more room for the 
generation of a large scale magnetic field through inverse cascade. 
Simulations were performed for $\epsilon=0$, $0.1$, and $0.2$, 
corresponding respectively to the MHD case, $k_{Hall}=10$ and $k_{Hall}=5$.\

\begin{figure}
\plotone{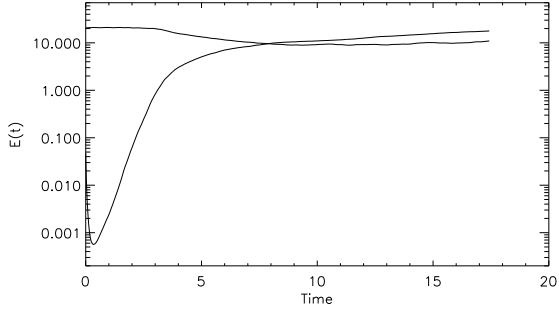}
\caption{Magnetic (below) and kinetic energy (above) as a function 
         of time for $ \epsilon = 0 $ and $k_{force}=10$.
\label{mhdsm_ener}}
\end{figure}

\begin{figure}
\plotone{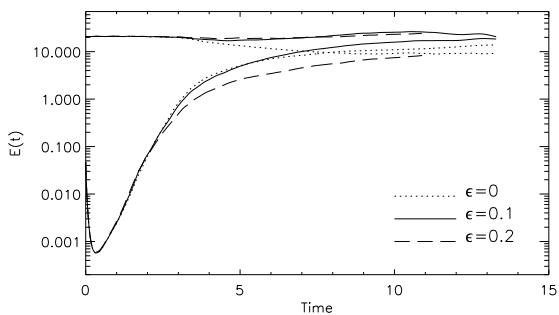}
\caption{Magnetic (below) and kinetic energy (above) as a function 
         of time for two runs with $ \epsilon = 0.2$, $0.1$ and 0 
         ($k_{force}=10$).
\label{hallsm_ener}}
\end{figure}

Figure \ref{mhdsm_ener} shows the evolution of the magnetic and kinetic 
energy in the MHD simulation. After saturation, a large scale magnetic 
field grows on a resistive time scale, as will be shown in the energy 
spectrum and was previously observed in large scale dynamo simulations 
\citep{Brandenburg}. Note that the system finally reaches a state of 
super-equipartition, i.e. a level of magnetic energy which is larger than 
the kinetic energy. Figure 
\ref{hallsm_ener} shows its counterpart when $\epsilon=0.1$ and $0.2$. 
In the simulation with $\epsilon=0.1$, the growth of magnetic energy 
after the saturation is clearly faster, and the system reaches a final 
state with more magnetic energy than in the MHD run.

\begin{figure}
\plotone{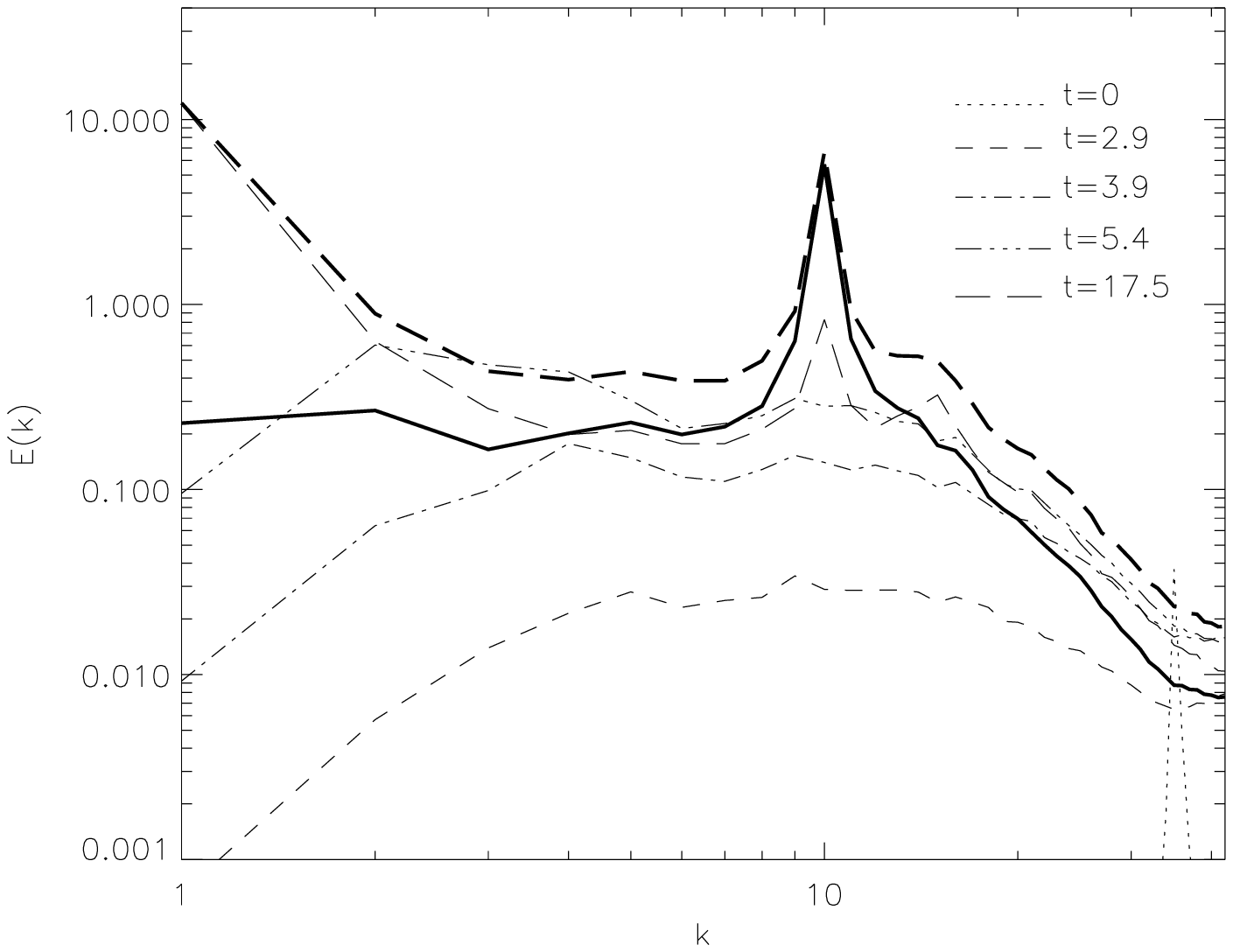}
\caption{Mean kinetic energy spectrum (thick line), total energy 
         spectrum (thick dashed line), and magnetic energy spectrum 
         at different times ($\epsilon = 0$ and $k_{force}=10$).
\label{mhdsm_spec}}
\end{figure}

\begin{figure}
\plotone{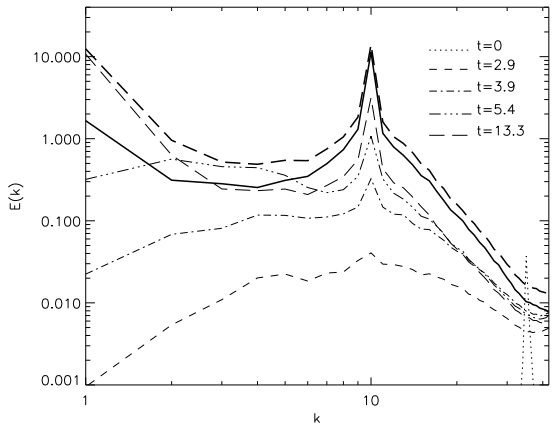}
\caption{Mean kinetic energy spectrum (thick line), total energy 
         spectrum (thick dashed line), and magnetic energy spectrum 
         at different times ($\epsilon = 0.1$ and $k_{force}=10$).
\label{hallsm_spec}}
\end{figure}

Figures \ref{mhdsm_spec} and \ref{hallsm_spec} show the 
energy spectrum at different times for runs with $\epsilon=0$ and 
$\epsilon=0.1$, respectively. Note that in both simulations, 
the magnetic energy at intermediate scales ($2<k<8$) starts to decay
after saturation ($t>3.5$), while magnetic energy at the largest scale ($k=1$) keeps 
growing. As a result, after $t=8$ the system reaches the 
state of super-equipartition in the MHD case. This is even more clear in Fourier space 
(Figure \ref{mhdsm_spec}), where magnetic energy in the shell $k=1$ is 
two orders of magnitude larger than kinetic energy at $t=17.5$. An 
excess of magnetic energy can also be observed at small scales. 

In the Hall-MHD case, we observe again a faster growth of the large 
scale magnetic field (Figure \ref{hallsm_spec}), but with a final state with 
super-equipartition only at large scales. Moreover, the kinetic energy in the 
shell $k=1$ is one order of magnitude larger than in the MHD case, and at 
small scales we obtain sub-equipartition between magnetic and kinetic energy. 

\begin{figure}
\plotone{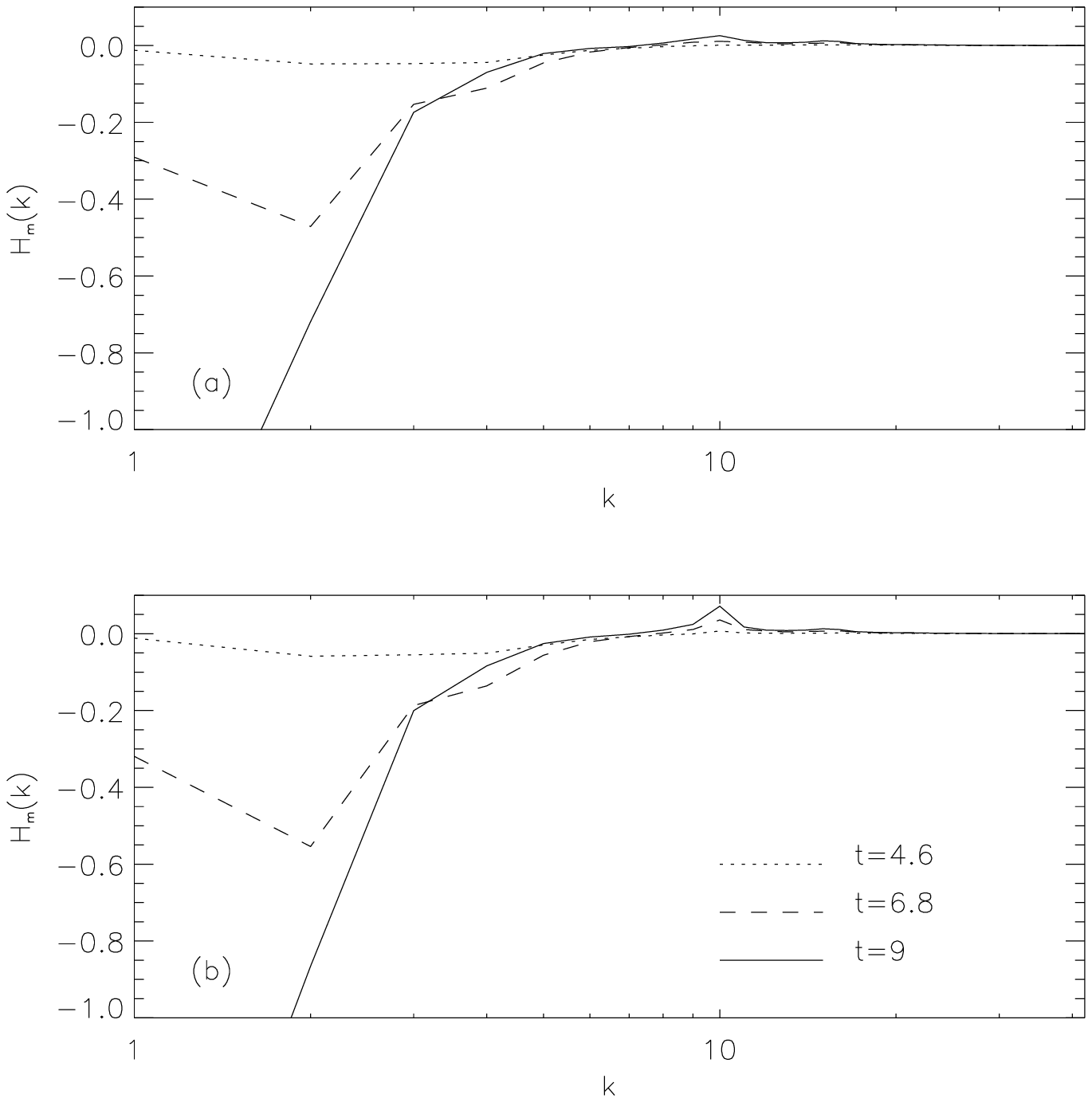}
\caption{Magnetic helicity spectra for: (a) $ \epsilon = 0$, and (b) 
         $ \epsilon = 0.2$ ($R=300$).
\label{helispec}}
\end{figure}

Figure \ref{helispec} shows the magnetic helicity spectrum as a function 
of time for different values of $\epsilon$. As mentioned in Section 4, 
the $ \alpha $ effect creates magnetic helicity of a sign opposite to that of 
kinetic helicity at large scales. This effect is balanced by the creation 
of an opposite amount of magnetic helicity at small scales. Therefore, the
diffusion preferentially destroys the short-scale magnetic helicity in 
reconnection events, leaving a net helicity of opposite sign at large 
scales \citep{Brandenburg}. \citet{MGM3} suggested that while the Hall-MHD 
dynamo process also creates equal and opposite amounts of magnetic helicity 
at large and at small scales, the dissipation of magnetic helicity at 
small scales is less efficient as $\epsilon$ is increased. Figure 
\ref{helispec} shows that when the Hall effect is present, even at late 
times an excess of positive magnetic helicity at small scales ($k\ge10$) 
can  be readily identified in the spectra.

\section{DISCUSSION}

In this paper we have  presented the  results of direct numerical 
simulations of turbulent dynamo action in Hall-MHD. We find that
with increasing Reynolds number and scale separation, the  
Hall MHD  dynamo works more efficiently when the Hall length is close 
but larger than the dissipation scale (Hall-enhanced regime). For larger 
values of $\epsilon$ the Hall MHD dynamo is less efficient. In addition, 
the value of $\epsilon$ (which measures the strength of the Hall term) 
at which the dynamo is most efficient decreases at higher Reynolds numbers.  

An acceleration of the process responsible for the growth of a large 
scale magnetic field is observed at moderate values of $\epsilon$. 
Although these simulations are made at Reynolds numbers which are 
far away from realistic values for astrophysical plasmas, the  
results obtained are encouraging; the dynamos tend to work better at high 
Reynolds numbers

By calculating the magnitude and nature of the generated magnetic 
field as the amplitude of the Hall term is varied, we obtain new 
evidence showing that the Hall dynamo can be fundamentally different 
from its classical MHD counterpart.

\acknowledgements

Use of the Beowulf cluster {\it Bocha} at Departamento de F\'{\i}sica,
FCEN, UBA is acknowledged. The authors gratefully acknowledge A. Brandenburg and 
A. Pouquet for very fruitful and enlightening comments while reviewing the 
manuscript, and to the Abdus Salam 
International Centre for Theoretical Physics, were the initial stages 
of this study were performed. Research of SMM was supported by US DOE 
contract DE-FG03-96ER-54366. Research of DOG and PDM has been partially funded 
by grant X209/01 from the University of Buenos Aires and by grant PICT 03-9483 
from ANPCyT. PDM is a fellow of CONICET, and DOG is a member of the Carrera del 
Investigador Cientifico of CONICET.

\end{document}